\documentclass[]{aastex631}
\usepackage{amsmath}
\usepackage{graphicx}
\usepackage{threeparttable}
\usepackage{txfonts}
\usepackage{soul}
\usepackage{tabularx} 

\shorttitle{SF Revealing the Origin of FRBs}
\shortauthors{Li et al.}

\begin{document}

\title{Structure Functions of Rotation Measures Revealing the Origin of Fast Radio Bursts}

\correspondingauthor{Fa-Yin Wang}
\email{fayinwang@nju.edu.cn}

\author[0009-0007-3326-7827]{Rui-Nan Li}
\affiliation{School of Astronomy and Space Science, Nanjing University Nanjing 210023, People's Republic of China}
\author[0000-0002-2171-9861]{Zhen-Yin Zhao}
\affiliation{School of Astronomy and Space Science, Nanjing University Nanjing 210023, People's Republic of China}

\author[0000-0001-6021-5933]{Qin Wu}
\affiliation{School of Astronomy and Space Science, Nanjing University Nanjing 210023, People's Republic of China}

\author[0000-0003-0672-5646]{Shuang-Xi Yi}
\affiliation{School of Physics and Physical Engineering, Qufu Normal University, Qufu 273165, People's Republic of China}

\author[0000-0003-4157-7714]{Fa-Yin Wang}
\affiliation{School of Astronomy and Space Science, Nanjing University Nanjing 210023, People's Republic of China}
\affiliation{Key Laboratory of Modern Astronomy and Astrophysics (Nanjing University) Ministry of Education, People's Republic of
	China}

\begin{abstract}

The structure function (SF) analysis is a powerful tool for studying plasma turbulence. Theoretically, the SF of Faraday rotation measure (RM) is expected to include a geometric component due to the relative orientation of sightlines through an ordered magnetic field. However, observational evidence for this component remains elusive. Here, we report that the SFs of the binary PSR B1744-24A and the repeating fast radio burst (FRB) 20201124A exhibit both a periodic geometric component, caused by binary orbital motion, and a flat statistical component. The statistical component, induced by stochastic fluctuations in electron density and magnetic field, aligns with RM scatter derived from pulse depolarization. These findings indicate that FRB 20201124A has a binary origin and suggest that the periodic geometric component can serve as a diagnostic tool to identify binary companions.
\end{abstract}


\section{Introduction} \label{sec:intro}
Fast radio bursts (FRBs) are luminous \citep{Lorimer2007}, millisecond-duration radio flashes originating from extragalactic sources with unknown physical origins \citep{Cordes19,Zhang2022,Wu2024}. Currently, advancements in highly sensitive radio telescopes such as the Five-hundred-meter Aperture Spherical radio Telescope (FAST) and the Green Bank Telescope (GBT) have facilitated comprehensive polarization measurements for fast radio bursts (FRBs) and pulsars. Notably, active repeaters like FRB 20201124A, FRB 20220912A, and the binary system PSR B1744–24A have been extensively observed for their polarization properties \citep{Xu2022,Zhang2023,Li2023,Feng2024}. Some observations suggest that some repeating FRBs exhibit polarization behaviors similar to those of binary systems \citep{Wang2022,Li2023,Anna-Thomas2023,Zhao2023,Rajwade2023}, providing critical insights into the local environments of their sources.

The rotation measures (RMs) of repeating FRBs are particularly valuable as they allow tracking of temporal changes over periods ranging from short \citep{Luo2020,Hilmarsson2021} to long intervals \citep{Hilmarsson121102,Xu2022}. Observations indicate that significant RM variations are a hallmark of repeating FRB sources \citep{Mckinven2023}, suggesting the presence of dynamic magneto-ionic environments. These environments may include the wind nebulae of young magnetars \citep{86,Yang2019,Zhao2021}, supernova remnants \citep{Piro2016,Yang2017,Piro2018,Zhao2021a}, the stellar wind and disks of companions \citep{Wang2022,Anna-Thomas2023,Zhao2023}, AGN disks \citep{Zhao2024} or outflows from massive black holes \citep{90,88,LiR2023}. \cite{Yang2023} has investigated some astrophysical processes that may cause RM variations of FRBs. Despite these expectations, direct evidence pinpointing the exact origins of these dynamic magnetized environments remains elusive.

Investigations of turbulence, which is prevalent in astrophysical plasmas \citep{Biskamp2003,Elmegreen2004}, could be key to distinguishing these environments. Turbulence is known to occur in various settings, including the solar wind, the interstellar and intergalactic mediums, accretion disks, and supernova remnants, playing a pivotal role in numerous astrophysical processes. These include the amplification of cosmic magnetic fields \citep{Ryu2008}, star formation \citep{McKee2007}, particle acceleration \citep{Zhang2011,Lazarian2020}, and magnetic reconnection \citep{Wang2023}. Turbulence naturally causes fluctuations in density and magnetic fields, which in turn lead to variations in RMs. The analysis of RMs using the Structure Function (SF) is a crucial statistical method for studying turbulence \citep{Kolmogorov1941,Simonetti1984,Simonetti1986,Minter1996,Clegg1992,Lazarian2016,Xu2016}. As a result, SF analysis is also a vital tool for probing the plasma environments surrounding FRBs. 

In general, the SF of RM consists of a statistical component and a geometric component \citep{Simonetti1984,Clegg1992}. The stochastic fluctuations in free electron density and magnetic field will induce the statistical component. At large spatial separations, the SF of a source traveling in an extended homogeneous medium with an ordered magnetic field will have a substantial geometric component \citep{Simonetti1984,Clegg1992}, due to the changes in the line of sight (LOS) relative to the magnetic field direction. 
In order to properly understand any observational results, we must distinguish between the statistical and geometric components.
However, clear evidence of the geometric component has not yet been established from observations. In binary systems, the quasi-ordered magnetic field in the stellar wind or the decretion disk of the companion star could produce a geometric component in RM SFs, as shown in Figure \ref{fig1}. This geometric component will show periodic behavior due to the orbital motion. So, this characteristic feature can be used to examine the binary model for RM variations.

In this study, we aimed to investigate the properties of plasma turbulence near sources of FRBs by employing SF analysis. Additionally, we sought to determine the presence of the geometric component within FRB and binary system. A detailed description of the data utilized for this analysis and an introductory overview are provided in Section \ref{sec:SF}. The results of SF analysis of RM and DM for the spider system PSR B1744–24A and seven active repeating FRBs are performed in Section \ref{sec:results}. Discussion and conclusions are presented in Section \ref{sec:dis}.

\section{Datasets and Structure Function} \label{sec:SF}
\subsection{Datasets}
The datasets used in structure function analysis are summarized in Table \ref{tab:data}. We use the bursts of FRB 20180916B with both RM and DM measurements by Canadian Hydrogen Intensity Mapping Experiment (CHIME) \citep{Mckinven2023,CHIME2019,CHIME2024}, Low-Frequency Array (LOFAR) \citep{Pleunis2021,Gopinath2024} and the upgraded Giant Metrewave Radio Telescope (uGMRT) \citep{Bethapudi2024}. The RM and DM data of FRB 20201124A from Five-hundred-meter Aperture Spherical radio
Telescope (FAST) \citep{Xu2022,Jiang2022,Feng2022} is used. FRB 20190520B has few bursts with RM and DM measurements \citep{Anna-Thomas2023,Feng2022,Niu2022}. The number of DM data of FRB 20121102A is large. Meanwhile, the RM data is also adequate for analysis \citep{Spitler2016,Hewitt2022,Michilli2018,Hilmarsson121102,Li2021,Gajjar2018}. There are a limited number of RM and DM data of FRB 20190303 for SF analysis \citep{Mckinven2023,Feng2022,Fonseca2020,CHIME2024}. FRB 20220912A has more than 1,000 bursts detected by FAST \citep{Zhang2023,Feng2024}. We adopt the RM and DM data of PSR B1744-24A observed by GBT \citep{Li2023}.

\begin{table*}
\centering
\caption{The Data Used for Structure Function Analysis}
\label{tab:data}
\scriptsize
\setlength{\tabcolsep}{2pt}
\renewcommand{\arraystretch}{1.1}
\begin{tabularx}{\textwidth}{X c c X}
\hline 
Source & Number of RM observation & Number of DM observation & Refs. \\
\hline 
PSR B1744-24A & 625 & 596 & \cite{Li2023} \\
FRB 20220912A & 1204 & 1076 & \cite{Zhang2023,Feng2024} \\
FRB 20201124A & 1679 & 811 & \cite{Xu2022,Jiang2022,Feng2022,Hilmarsson2021} \\
FRB 20190520B & 16 & 207 & \cite{Anna-Thomas2023,Feng2022,Niu2022} \\
FRB 20190303A & 17 & 55 & \cite{Mckinven2023,Feng2022,Fonseca2020,CHIME2024} \\
FRB 20180916B & 74 & 54 & \cite{Mckinven2023,Gopinath2024,Pleunis2021,CHIME2019,Chawla2020,CHIME2024,Bethapudi2024} \\
FRB 20180301A & 20 & 61 & \cite{Kumar2023,Luo2020,Feng2022} \\
FRB 20121102A & 43 & 1724 & \cite{Spitler2016,Hewitt2022,Michilli2018,Hilmarsson121102,Li2021,Gajjar2018} \\
\hline
\end{tabularx}
\end{table*}

\subsection{Structure function} \label{subsec:sf}
Considering an observational quantity $A(x)$ which is a function of position or time, $x$ denotes the position on the plane of sky or the time of observation. 
The structure function (SF) represents the mean-square difference between two observational quantities with spatial separation or temporal separation $\Delta x$. SF is defined as
\begin{equation}
D_{\mathrm{A}}({\Delta x}) \equiv\left\langle[\mathrm{A}({x}+\Delta x)-\operatorname{A}({x})]^2\right\rangle,
\end{equation}
where $\left\langle...\right\rangle$ represents an ensemble average. 
We neglect the distance along LOS of different observational points. Based on the statistical descriptions \citep{Lazarian2016}, the power-law model of SF is sufficient to capture the scaling properties of turbulence
\begin{equation} \label{equ:1}
\widetilde{D}_A({\Delta x})=2 \sigma_A^2 ~\frac{\left(\Delta x\right)^{m_A}}{l_A^{m_A}+\left(\Delta x\right)^{m_A}},
\end{equation}
where the variance of fluctuations is defined as
\begin{equation}
\sigma_A^2=0.81^2\left\langle\delta\left(n_e B_{\|}\right)^2\right\rangle
\end{equation}
for RM density and 
\begin{equation}
\sigma_A^2=\left\langle\delta\left(n_e\right)^2\right\rangle
\end{equation} \label{equation:spe}
for DM density. $l_A$ is the correlation scale of density fluctuations which distinguishes the energy dissipation scale and energy injection scale. Scales above the inner scale $l_A$ of shallow spectrum and scales below the outer scale $l_A$ of the steep spectrum are considered as the inertial range where equation (\ref{equ:1}) is applicable. $m_A$ is the index of power-law functions. 
The index of SF can be used to determine the spectral index $\alpha$ which depends on the turbulent spectrum is whether shallow or steep \citep{Lazarian2016},
\begin{equation}
\begin{gathered}
\alpha=m_\phi-N, \quad \alpha>-3, \\
\alpha=-m_\phi-N, \quad \alpha<-3,
\end{gathered}
\end{equation}
where $N$ is the dimensionality of turbulence in space. 
The theory of Kolmogorov turbulence was widely used in the research on IGM and ISM. The Kolmogorov turbulence theory assumes the energy of large eddies was cascaded to the small eddies by kinetic-energy-conserving interactions where it was dissipated \citep{Elmegreen2004}. The Kolmogorov turbulence can be considered to be locally homogeneous and isotropic in hydrodynamic case. MHD turbulence exhibits more complexity compared to hydrodynamic case. In magnetohydrodynamics, the magnetic field establishes a preferred direction \citep{Montgomery1981,Cho2002, Higdon1984} leading to anisotropic statistical properties in the turbulence when it is affected by magnetization \citep{Lazarian2016, Lazarian1999}. This kind of turbulence was found in different astrophysical environments. The statistical analysis reveals the presence of Kolmogorov spectra in Saturn's magnetosphere \citep{Xu2023}, Jupiter's magneto sheath \citep{Bandyopadhyay2021}, solar wind plasma \citep{Shaikh2010}, star winds interacting with ISM \citep{Garcia2020} and specific scale of ISM \citep{Minter1996,Xu2020}.

The measurements of RM and DM provides us with unique information on the magnetized turbulence in the local environment. DM is defined as
\begin{equation}
\mathrm{DM}= \int_0^L n_e(l) \mathrm{d} l,
\end{equation}
where $n_{e}$ is the number density of free electrons. 
SFs of RM and DM depend on the power spectra $\mathcal{P}(k)$ of the fluctuations in $\left(n_e B_{\|}\right)(\vec{x})$ and $\left(n_e \right)(\vec{x})$, where $k$ is the spatial wavenumber. Here we assume the power spectra of RM and DM follow a power law with slope $\alpha$ ($\alpha$ being negative), i.e., $\mathcal{P}(k) \sim k^{\alpha}$. The screen can be thin when the correlation length of turbulent fluctuations exceeds the line-of-sight extent of the Faraday screen: $\Delta R<l_{\mathrm{RM}}$. The SF of RM can be expressed as \citep{Lazarian2016,Xu2016}

\begin{equation} \label{equ:spectral_index}
D_{\mathrm{RM}}(\Delta x) \sim 
\left\{
\begin{array}{ll}
\sigma_{\mathrm{RM}}^2 \Delta R \left( \frac{\Delta x}{l_{\mathrm{RM}}} \right)^{-(\alpha+2)}, & \Delta x < L \\
\sigma_{\mathrm{RM}}^2 \Delta R^2 \left( \frac{\Delta x}{l_{\mathrm{RM}}} \right)^{-(\alpha+3)}, & \Delta R < \Delta x < l_{\mathrm{RM}} \\
\sigma_{\mathrm{RM}}^2 \Delta R^2, & \Delta x > l_{\mathrm{RM}} \\
\end{array}
\right.
\end{equation}

For above equations, $L\sim \Delta R$, $\sigma_{\rm{RM}} = \kappa^2\left\langle\delta\left(n_e B_{\|}\right)^2\right\rangle$ and $\sigma_{\rm{DM}} = \kappa^2\left\langle\delta\left(n_e\right)^2\right\rangle$. The SF of DM is similar to the expressions above. It can be inferred that $D_{\mathrm{RM}}(\Delta x) \propto l^{-(\alpha + 2)}$ or $D_{\mathrm{RM}}(\Delta x) \propto l^{-(\alpha + 3)}$ in the inertial range, and $D_{\mathrm{RM}}(l) \sim $ constant beyond the inertial range. Nevertheless, there is a limited effective range of the relation between SF and power spectrum slope, and the slope of the power spectrum may be steeper than that in this range. 

When we calculate the SFs of RM and DM, each pair of RM and DM need to be corrected by subtracting ($\sigma_{\rm{RM}}^{2}(x)+\sigma_{\rm{RM}}^{2}(x+\Delta x)$) and ($\sigma_{\rm{DM}}^{2}(x)+\sigma_{\rm{DM}}^{2}(x+\Delta x)$) to remove the noise bias caused by measurement uncertainties. 

Three-dimensional (3-D) Kolmogorov turbulence has a value of $\alpha = -11/3$ \citep{Kolmogorov1941}, which corresponds to the steep spectrum, $D_{\mathrm{RM}}(l) \propto l^{5/3}$ or $D_{\mathrm{RM}}(l) \propto l^{2/3}$. When the transverse scale is much larger than the thickness of the observed Kolmogorov turbulent region, the power-law index is $2/3$ which reflects the behavior of 3-D turbulence measured in a 2-D geometry. It is important to note that this is distinct from purely 2-D turbulence, which exhibits an inverse cascade and different spectral properties. Table \ref{tab:PW} summarizes the power-law relations for Kolmogorov turbulence including the two-dimensional distribution of the fluctuating field. The temporal SF that is converted from the spatial SF by assuming a constant velocity of the source.

\begin{table}[ht]
\centering
\caption{Power-law Relations for Turbulence Medium}
\label{tab:PW}
\begin{tabular}{ccccc}
\hline &  &\multicolumn{2}{c}{ Exponent } \\
\hline Quantity & & \begin{tabular}{l}
$3-\mathrm{D}$ \\

\end{tabular} & \begin{tabular}{l}
$2-\mathrm{D}$ \\

\end{tabular} \\
\hline 2-D, 3-D power spectrum & $\mathcal{S}_2\left(q_s\right), \mathcal{S}\left(q_s\right)$ & $\alpha = -11 / 3$ & $\alpha = -11 / 3$ \\
\hline Structure function & $D_\phi(d)$  & $-(\alpha+2) = 5 / 3$ & $-(\alpha+3) = 2 / 3$ \\
\hline Temporal structure function & $D_\tau(\tau)$ & $-(\alpha+2) = 5 / 3$ & $-(\alpha+3) = 2 / 3$ \\
\hline
\end{tabular}
\tablecomments{Power-law relations of temporal structure function are only valid under proper motion assumption.}
\end{table} 

\section{Results} \label{sec:results} 

A SF measures the amount of fluctuations in a quantity as
a function of the time of the fluctuations. The temporal RM SF can be described as \citep{Simonetti1984,Mckinven2023} 
\begin{equation}
D_{\mathrm{RM}}(\Delta t)= \left \langle [\mathrm{RM}(t)-\mathrm{RM}(t+\Delta t)]^2 \right \rangle,
\end{equation}
where the angle brackets denote an ensemble average over time separation $\Delta t$. RM is defined as
\begin{equation}
\mathrm{RM}=8.1 \times 10^{5} \int_{\operatorname{LOS}} n_{\mathrm{e}} \boldsymbol{B} \cdot \mathrm{d} \boldsymbol{l}~\rm{rad~m^{-2}},
\end{equation}
with the electron number density $n_{\mathrm{e}}$ (in units of cm$^{-3}$), the magnetic field $\boldsymbol{B}$ (in units of G) and the distance $l$ (in units of pc).

We first apply the SF analysis to the binary system PSR B1744-24A with uncommon polarization behaviors. The result is shown in panel (a) of Figure \ref{fig2}. The orange plus signs represent the measurements of $D_{\mathrm{RM}}$, and the blue points are the rebinned data. We can see that the SF of PSR B1744-24A shows a stochastic fluctuation and remains nearly constant for $\log(\Delta t)<-2$ day. For $\log(\Delta t)>-2$ day, it exhibits a trend that is similar to the trigonometric function and then enters into a decline phase, which is hard to explain by the turbulence theory \citep{Elmegreen2004}. The spatial SFs of the interstellar medium (ISM) \citep{Minter1996,Clegg1992,Haverkorn2008}, intergalactic medium (IGM) \citep{Xu2020,Xu2021} and supernova remnants (SNRs) \citep{Shimoda2018,Saha2019} show rising power-law or flat form depending on the physical scale, instead of a decline phase as found in PSR B1744-24A. The RM variation has been interpreted as the orbital motion leading to the changes of the angle between LOS and the magnetic field in the stellar wind \citep{Li2023}. We will try to explain it using the geometric component. 

\begin{figure} 
\centering
\includegraphics[width=140 mm]{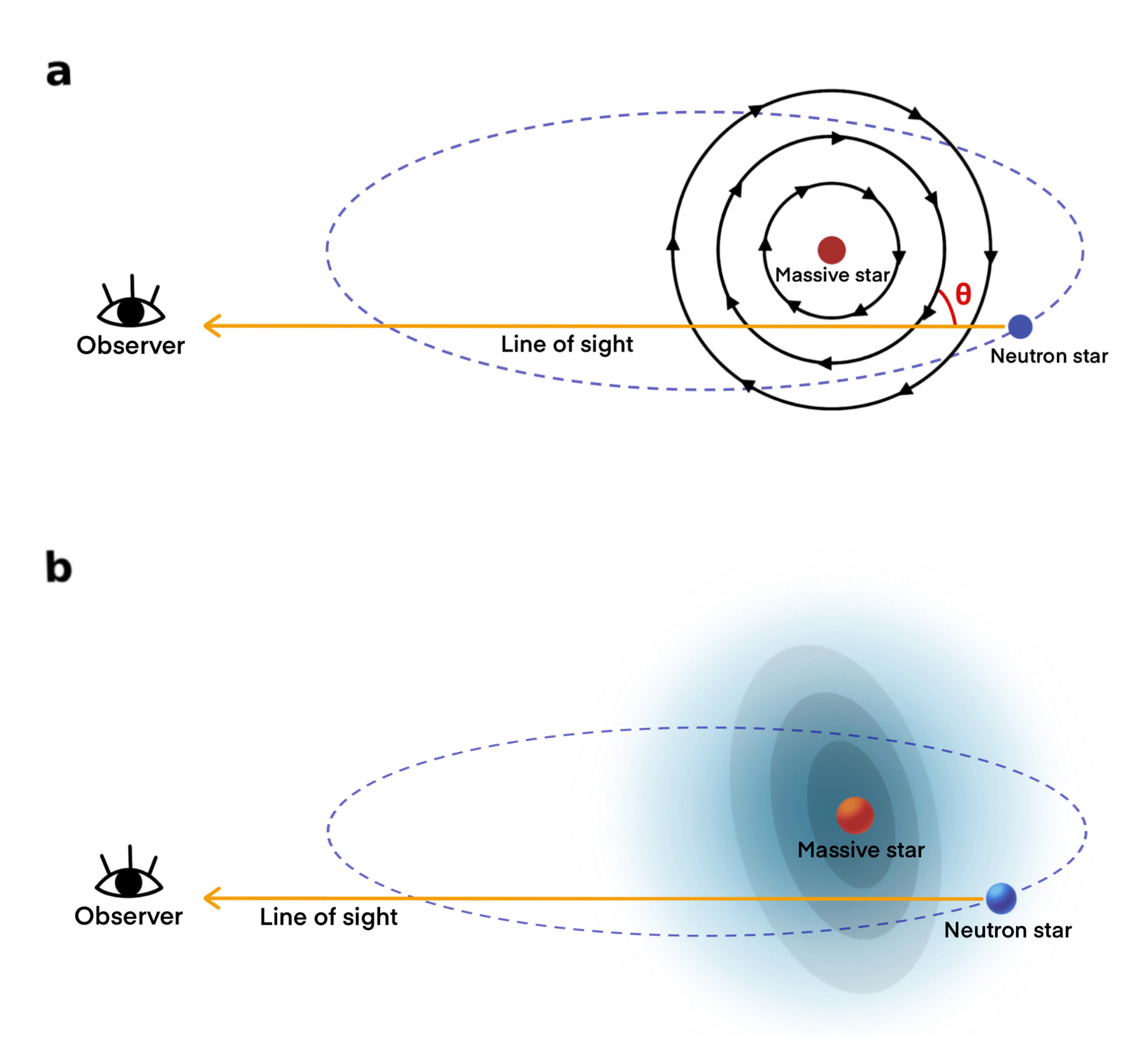}
\caption{The schematic diagram of binary systems. Panel (a): The top-down view of binary systems. The black solid lines with arrow correspond to the toroidal magnetic field line in the decretion disk or winds. The yellow line is the line of sight (LOS). The changes in the angle between the LOS and the magnetic field contribute to the geometric component in RM SF. 
Panel (b): The three-dimensional schematic diagram of binary systems. The grey shaded area indicates decretion disk of the companion star. The blue shaded area indicates stellar wind of the companion star.}
\label{fig1}
\end{figure}

\begin{figure} 
\centering
\includegraphics[width=180 mm]{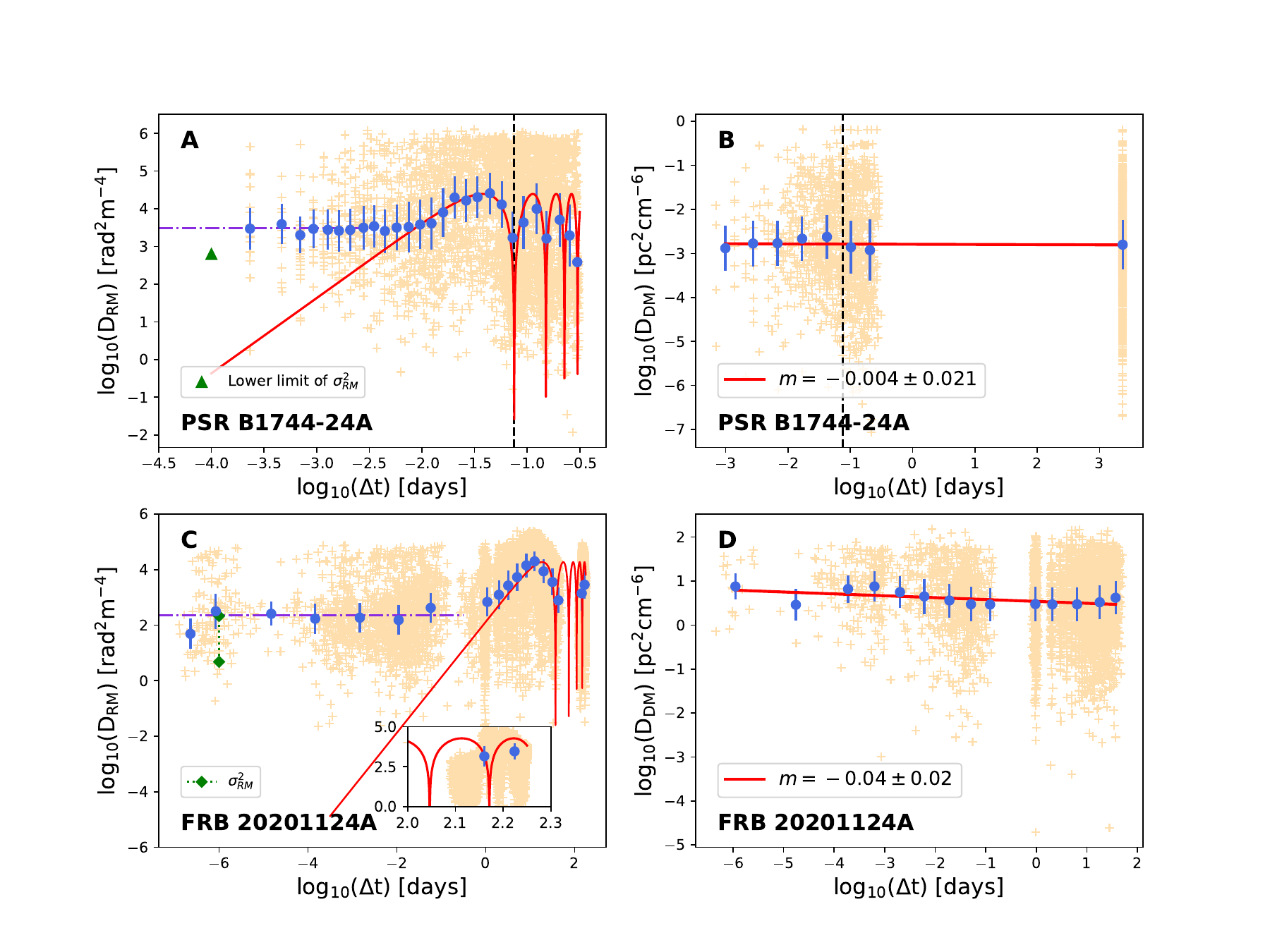}
\caption{The SFs for PSR B1744-24A and FRB 20201124A.
panel (a): SF of RM for PSR B1744-24A. Orange plus symbols represent the non-binned data after performing correction, and blue circles indicate binned data with uncertainties estimated as the standard error on the mean. The red solid line is the fit using equation (\ref{equ:2}) with $\rm{RM_0}=157\pm40~\rm{rad~m^{-2}}$ and the orbital period is $\sim1.8$ hours. The dot solid line corresponds to the statistical component $D_s=2927.14~\rm{rad^{2}~m^{-4}}$. The black dashed line corresponds to the orbital period of $1.8$ h derived in previous research \citep{Li2023}. 
panel (b): SF of DM for PSR B1744-24A. Red solid line corresponds to linear fit with power-law index $-0.004\pm0.021$. 
panel (c): SF of RM for FRB 20201124A. $\rm{RM_0}=136\pm24~\rm{rad~m^{-2}}$ and the orbital period $P=37\pm1$ days. The statistical component is $D_s=221.88~\rm{rad^{2}~m^{-4}}$.  
panel (d): SF of DM for FRB 20201124A. The power-law index is $-0.04\pm0.02$. 
}
\label{fig2}
\end{figure}

\begin{figure} 
\centering
\includegraphics[width=160 mm]{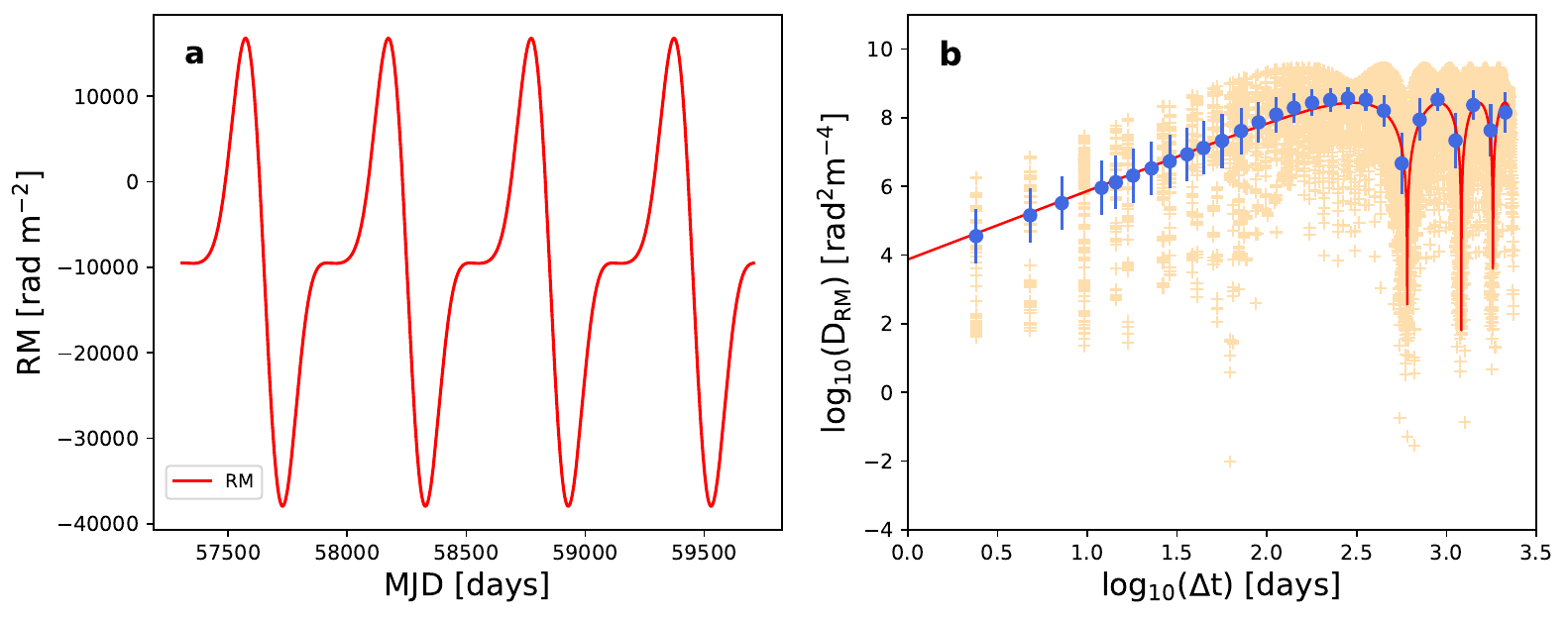}
\caption{Test the geometric component using simulations.  Panel (a): Simulated RM of FRB 20190520B. The red line represents the RM data simulated by the binary model. The model parameters are the same as those used for the observed RM fit \citep{Wang2022}. The period is taken as 600 days.
Panel (b): SF of simulated RM for FRB 20190520B. Orange plus symbols represent the non-binned data after doing correction, and blue circles indicate binned data with uncertainties estimated as the standard error on the mean. The red solid line is the fit using equation (\ref{equ:2}) with $\rm{RM_0}=16612\pm651~\rm{rad~m^{-2}}$ and the orbital period $P=605\pm3$ days which is consistent the one used in simulations \citep{Wang2022}.
}
\label{fig3}
\end{figure}

In a nearly homogeneous medium supported by the SF of DM, the geometric
component caused by the change in $\boldsymbol{B} \cdot d \boldsymbol{l}$ will dominate the RM SF \citep{Simonetti1984,Clegg1992}, as shown in Figure \ref{fig1}. For simplicity, we assume the angular velocity of the FRB source is a constant value $\omega$. In this case, the angle between the LOS and the magnetic field changes as $\Delta \theta = \rm{\omega}\Delta t$. The value of RM is RM$_0$ at an initial angle $\theta_0$. After some time $\Delta t$, the angle between the LOS and the magnetic field changes to $\theta = \theta_0 +\Delta \theta$. One has $\mathrm{RM}_0 \cos \left(\theta-\theta_0\right)$. The SF is computed by averaging RM differences across angular scales $\theta$ over a range of $0 \leq \theta \leq \pi$. The geometric contribution to the SF can be written as
\begin{equation} \label{equ:2}
D_g(\Delta \theta) = \left\langle[\rm R M(\theta)-R M(\theta+\Delta \theta)]^2\right\rangle=
\frac{1}{2}\mathrm{RM}_0^2(1-\cos \Delta \theta).
\end{equation}
As shown in the panel (a) of Figure \ref{fig2}, we fit the SF with equation (\ref{equ:2}) for $\log(\Delta t)>-2$ days, shown as red line.
The fitting result of average angular velocity is $\omega = 83.388\pm 3.468~\rm{rad~day^{-1}}$, which agrees well with the $1.8$-hour orbital period of this binary system. 
Thanks to the sufficient RM data across multiple periods, the periodic oscillation can be described by equation (\ref{equ:2}). The geometric component will show dips at $\omega \Delta t = 2n\pi$, where $n$ is an integer. The first dip occurs at $\Delta t$ equaling the orbital period $P$. We also test this speculation using the simulated RM data. The parameters of the binary model to explain the RM variation of FRB 20190520B are used \citep{Wang2022}. The orbital period is taken to be 600 days \citep{Wang2022}. The simulation lasts for 4 orbital periods to make the oscillation stage clear. The SF analysis for the simulated data is shown in Figure \ref{fig3}. The fit with equation (\ref{equ:2}) is shown as the red line. Because the effect of fluctuations in RM is not included in the simulation. We can see that the SF is completely dominated by the geometric component. The best fit for the period is $P=605\pm 3$ days, which is consistent with the period used in the simulation. This result proves that the RM SF in a binary system is dominated by the geometric component if RM variation is caused by the 
relative orientation of the lines of sight through an ordered magnetic field. 

The polarization behavior of some repeating FRBs is similar to PSR B1744-24A \citep{Li2023}. 
For comparison, we also calculate the SF of RM for FRB 20201124A, which is shown in panel (c) of Figure \ref{fig2}. 
The SF evolution is similar to that of PSR B1744-24A. They both show a trigonometric-like profile after a steady phase, followed by a decline which can not be interpreted by the turbulence theory. This feature can be well understood by the imprint of the geometric effect. As shown in the panel (c) of Figure \ref{fig2}, we also fit the SF with equation (\ref{equ:2}) for $\Delta t>1$ day, shown as the red line. The derived average angular velocity is $\omega=0.169\pm0.002~\rm{rad~day^{-1}}$. It is larger than the angular velocity corresponding to the 80-day orbital period derived from RM variation \citep{Wang2022}, i.e., $0.08~\rm{rad~day^{-1}}$. The reason is that equation (\ref{equ:2}) is only applicable for circular orbit. Moreover, a high eccentric orbit is required to explain the RM variation of FRB 20201124A \citep{Wang2022} and the angular velocity is relatively larger around the pericenter than that in the whole orbital phase.
Clear evidence of the geometric component has been found from the RM observations in PSR B1744-24A and FRB 20201124A for the first time.
The similarity between the SFs of RM for the two systems gives strong evidence that the source of FRB 20201124A is in a binary system. 

\begin{figure} 
\centering
\includegraphics[width=150 mm]{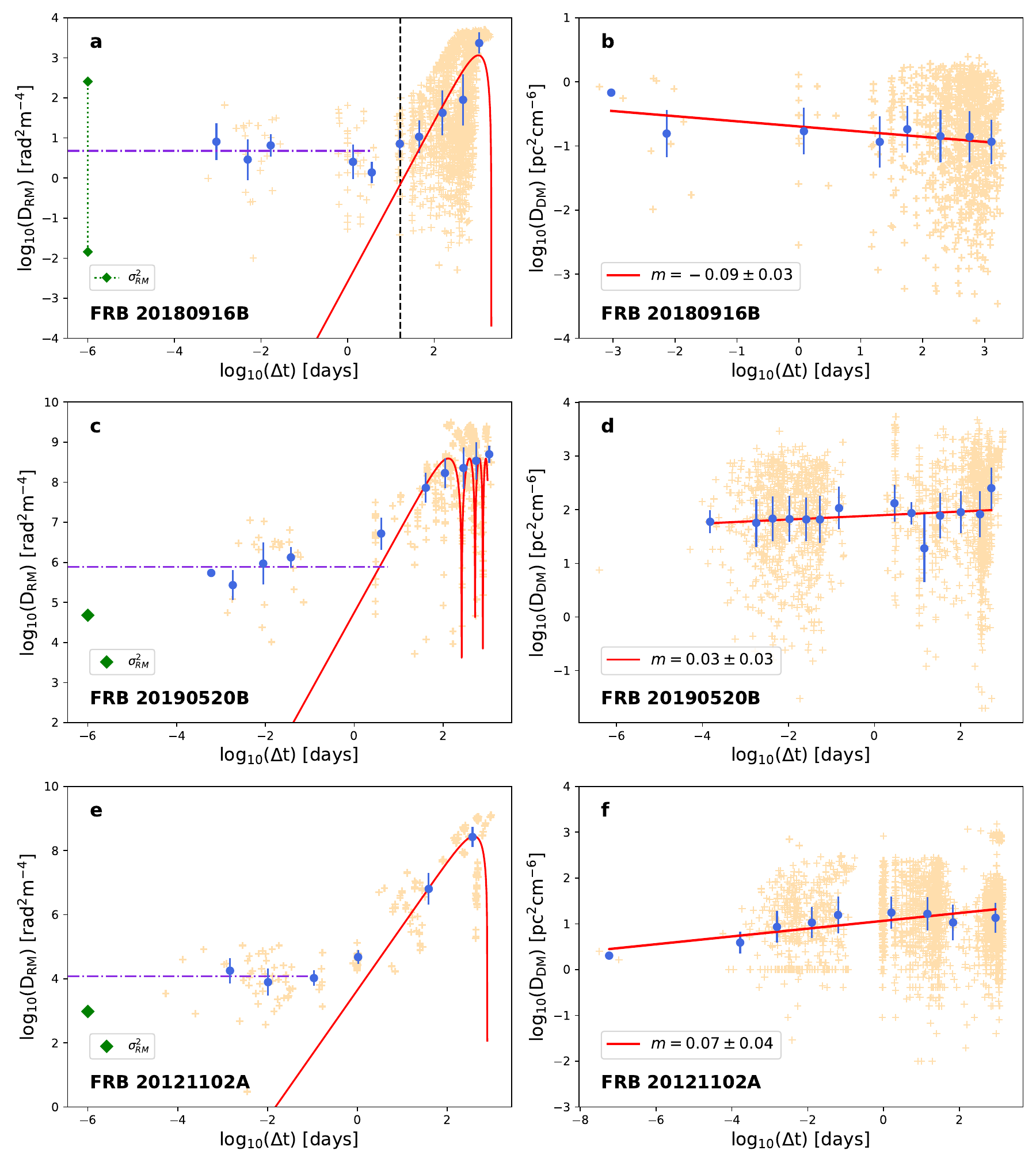}
\caption{The SFs for FRB 20180916B, FRB 20190520B, and FRB 20121102A. Panel (a): SF of RM for FRB 20180916B. Orange plus symbols represent the non-binned data after doing correction, and blue circles indicate binned data with uncertainties estimated as the standard error on the mean. The red solid line is the fit using equation (\ref{equ:2}) with $\rm{RM_0}>33~\rm{rad~m^{-2}}$ and the orbital period $P>2094$ days. The dot solid line corresponds to the statistic component $D_s=4.75~\rm{rad^{2}~m^{-4}}$. The black dashed line corresponds to the active period of $16.35$ days. 
Panel (b): SF of DM for FRB 20180916B. Red solid line corresponds to linear fit with power-law index $-0.09\pm0.03$. 
Panel (c): SF of RM for FRB 20190520B. $\rm{RM_0}>2\times10^4~\rm{rad~m^{-2}}$ and orbital period $\rm P > 524$ days and $D_s=7.6\times10^5~\rm{rad^{2}~m^{-4}}$. 
Panel (d): SF of DM for FRB 20190520B. The power-law index is $0.03\pm0.03$. 
Panel (e): SF of RM for FRB 20121102A. $\rm{RM_0}>1.7\times10^4~\rm{rad~m^{-2}}$ and the orbital period $P>785$ days and $D_s=1.2\times10^4~\rm{rad^{2}~m^{-4}}$. 
Panel (f): SF of DM for FRB 20121102A. The power-law index $-0.07\pm0.02$.
}
\label{fig4}
\end{figure}

We also perform the SF analysis for FRB 20180916B, FRB 20190520B, and FRB 20121102A. The results are shown in Figure \ref{fig4}. Similar to those of FRB 20201124A and PSR B1744-24A, the SFs of RM of three FRBs show a steady phase followed by an increasing phase. Considering the orbital motion in a binary system, we apply equation (\ref{equ:2}) to fit the increasing phase. Due to the lack of sufficient observational data and the inflection point shown in SF, the possible periodic oscillation in the RM SF contributed by the geometric component has not been found yet. The fitting results only indicate the lower limits of the orbital period. The potential orbital period could be much longer. Whether these three FRBs are in binary systems still needs to be tested by future observations. The fitting results are summarized in Table \ref{sffit}. 

The SFs of the above five sources all show a flat phase for small temporal separation. Below, we discuss its possible physical origin. Except for the geometric component, the statistical component $D_s(\Delta t)$ due to fluctuations in free electron density and magnetic field on time scales smaller than $\Delta t$ also contributes to the observed SF \citep{Simonetti1984,Clegg1992}. The total SF is
\begin{equation} \label{equ:3}
D(\Delta t)=D_g(\Delta t)+D_s(\Delta t).
\end{equation} 
If $\rm{RM_s}$ is the statistical contribution to $\mathrm{RM}$ from fluctuations on time scales
smaller than $\Delta t$, then $D_s=2\left\langle\mathrm{RM}_s^2\right\rangle$ \citep{Clegg1992}. 
The statistical components of SF are derived by calculating the mean value of $D_s$ after subtracting the geometric component $D_g$. The value of $D_s$ vary from $4 ~\rm{rad^2~m^{-4}}$ to $761409 ~\rm{rad^2~m^{-4}}$, as shown in Table \ref{sffit}. 

The statistical component of SF relates to the RM scatter $\sigma_{\rm{RM}}$ through $D_s\simeq \sigma_{\rm{RM}}^{2}$ \citep{Clegg1992}. 
The RM scatter term will cause pulse depolarization through multi-path propagation \citep{Feng2022,Wang2022}. Its value can be derived from fitting pulse depolarization. The values of $\sigma _{\rm{RM}}^2$ are $4.84 - 213.16~\rm{rad^2~m^{-4}}$ \citep{Lu2023}, $0.01-256~\rm{rad^2~m^{-4}}$ \citep{Mckinven2023b}, $47917.21~\rm{rad^2~m^{-4}}$ and $954.81~\rm{rad^2~m^{-4}}$ \citep{Feng2022} for FRB 20201124A, FRB 20180916B, FRB 20190520B and FRB 20121102A, respectively. Whereas, the range of $\sigma _{\rm{RM}}^2$ we used here is also considered unlikely for FRB 20180916B \citep{Mckinven2023b}. For PSR B1744-24A, the value of the statistical component is $38~\rm{rad~m^{-2}}$. From the observed pulse depolarization, the RM scatter ($\sigma_{\mathrm{RM}}$) is found to be larger than $25~\rm{rad~m^{-2}}$ \citep{Li2023}, which is in the range of $D_s$ value. From Table \ref{sffit}, the relation $D_s\simeq \sigma_{\rm{RM}}^{2}$ is valid for FRB 20201124A, PSR B1744-24A and FRB 20180916B with sufficient RM observations. 
The characteristic timescale for $\sigma_{\rm{RM}}$ is about $\Delta t\sim 1$ ms. Since RM variation has to occur
within the burst duration to cause depolarization. For all sources, the $D_s$ measurements can only be done at $\Delta t>10^{-6}$ days. For simplicity, a characteristic timescale of $\log(\Delta t) = -6~\rm{days}$ for $\rm{\sigma_{\rm{RM}}}$ has been chosen in Figure \ref{fig2} and Figure \ref{fig4}. Because the fit of $D_s$ is nearly constant, we extrapolate its value to $\Delta t\sim 1$ ms. The different timescales may cause the discrepancy between $D_s$ and $\rm{\sigma_{\rm{RM}}}$ for FRB 20190520B and FRB 20121102A. Additionally, RM measurements typically encompass measurement errors that are statistically independent, or nearly so. As a result, they may also introduce a flat component to the SF at small lags, with a level corresponding to twice the variance of these errors. This effect is coupled with the statistical effect. However, by removing the noise bias when calculating the SFs, the contributions from measurement errors has been eliminated.

\begin{figure*}
\centering
\includegraphics[width=160 mm]{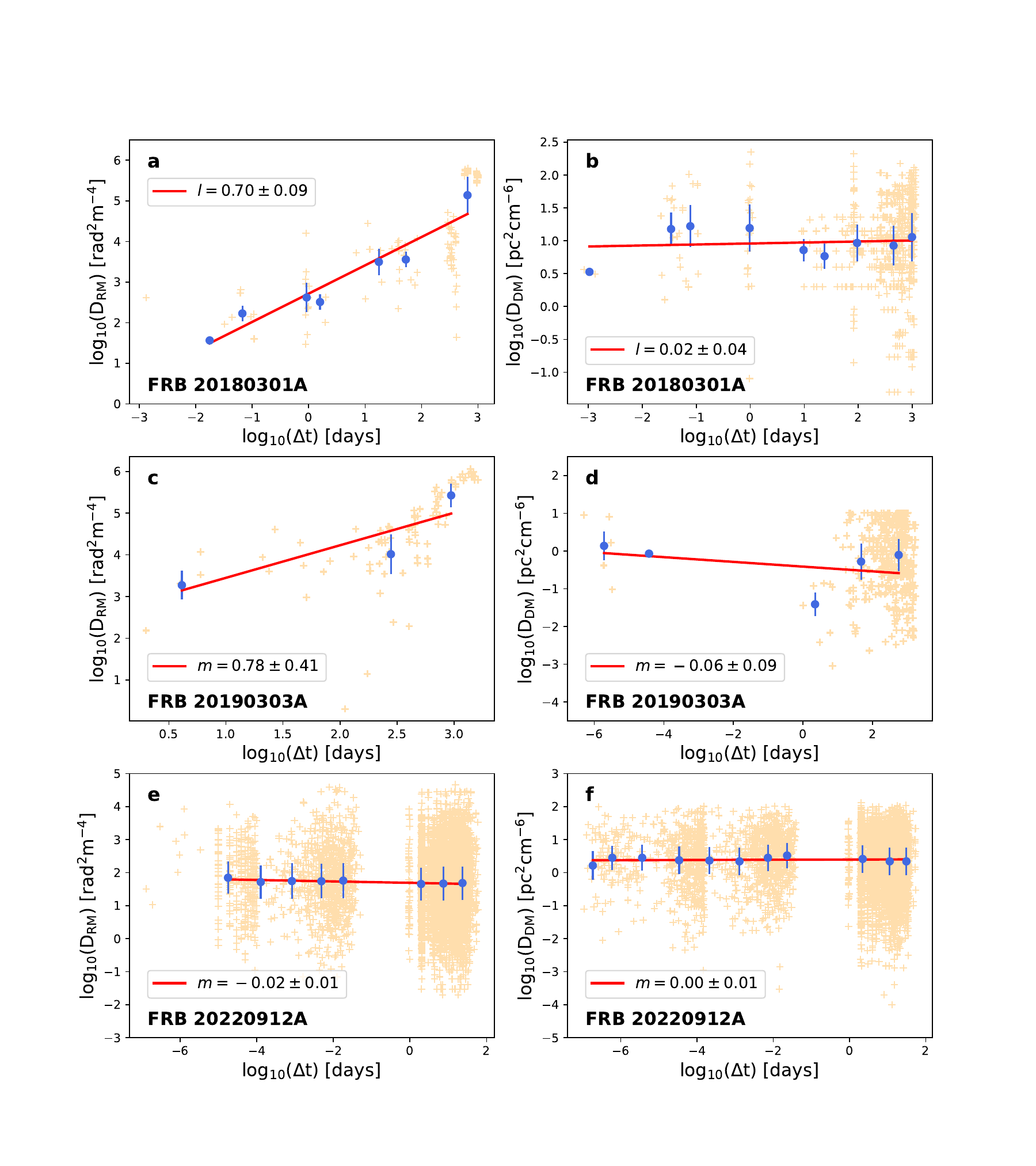}
\caption{The SFs for FRB 20180301A, FRB 20190303A, and FRB 20220912A. Panel (a): SF of RM for FRB 20180301A, Orange plus symbols represent the non-binned data after doing correction, and blue circles indicate binned data with uncertainties estimated as the standard error on the mean. The red solid line corresponds to linear fit with power-law index $0.70\pm0.09$. 
Panel (b): SF of DM for FRB 20180301A. Red solid line corresponds to linear fit with power-law index $0.02\pm0.04$. 
Panel (c): SF of RM for FRB 20190303A. Red solid line corresponds to linear fit with power-law index $0.78\pm0.41$. 
Panel (d): SF of DM for FRB 20190303A. Red solid line corresponds to linear fit with power-law index $-0.06\pm0.09$. 
Panel (e): SF of RM for FRB 20220912A. Red solid line corresponds to linear fit with power-law index $-0.02\pm0.01$. 
Panel (f): SF of DM for FRB 20220912A. Red solid line corresponds to linear fit with power-law index $0.00\pm0.01$. 
}
\label{Fig5}
\end{figure*}

We also study the RM SFs for FRB 20180301A, FRB 20220912A and FRB 20190303A. They are fitted with the power-law function ($D_{\rm{RM}}(\Delta t) \sim\Delta t^{m}$). The RM SFs for FRB 20180301A and FRB 20190303A exhibit a power-law form with an index around 0.7, shown in Figure \ref{Fig5}. If we assume the source travels at a constant velocity, i.e., the proper motion of a neutron star, the spatial separation is proportional to the temporal separation. So the spatial SF has the same behaviors as the temporal SF. The value is consistent with the expectation of the two-dimensional Kolmogorov turbulent medium ($m=2/3$). 
FRB 20220912A has small RM values and shows insignificant intraday RM changes with a mean value close to zero \citep{Zhang2023,Ravi2023}. This implies a clean environment around this source. The SF of RM is presented in Figure \ref{fig4}, which shows an extremely flat SF of RM. This suggests a non-evolving and uncorrelated RM.

\begin{table*} \label{sffit}
\centering
\caption{Fitting Results of Structure Functions}
\scriptsize 
\setlength{\tabcolsep}{2pt} 
\renewcommand{\arraystretch}{1.1} 
\begin{tabular}{cccccc}
\hline
\hline
{Source} & Angular & Orbital & {$RM_0$ (rad m$^{-2}$)} & {$D_s$ (rad m$^{-2}$)} & {$\sigma^2_{RM}$ (rad$^2$ m$^{-2}$)} \\
& velocity $\omega$ (rad day$^{-1}$) & period $P$ (day) & & & \\
\hline
PSR B1744-24A  & $83.388 \pm 3.468$ & $0.075 \pm 0.002$ & $157 \pm 40$ & $2927.14$ & $> 625$ \citep{Li2023} \\
FRB 20201124A  & $0.169 \pm 0.002$ & $37 \pm 1$        & $136 \pm 24$ & $221.88$  & $4.84 \pm 213.16$ \citep{Lu2023}\\
FRB 20180916B  & $<0.003$          & $>2094$           & $>33$        & $4.75$    & $0.01 \pm 0.25$ \citep{Mckinven2023b}\\
FRB 20190520B  & $<0.012$          & $>524$            & $>19799$     & $761409.86$ & $47917.21$ \citep{Feng2022}\\
FRB 20211202A  & $<0.008$          & $>785$            & $>16688$     & $12075.56$ & $954.81$ \citep{Feng2022}\\
FRB 20190520B (Simulated RMs) & $0.010 \pm 0.001$ & $605 \pm 3$  & $16612 \pm 651$ & /        & / \\
\hline
\end{tabular}
\tablecomments{The orbital period is calculated from the average angular velocity. The average angular velocity $83.388 \pm 3.468$ rad day$^{-1}$ of PSR B1744-24A agrees very well with its 1.8 h orbital period. The geometric component fitting results of FRB 20180916B, FRB 20190520B, and FRB 20211202A only indicate the lower limits of the potential orbital period and $RM_0$ due to a small sample of RM data and the absence of a turning feature in their SFs. The simulated RM data for FRB 20190520B are calculated by binary model.}
\end{table*}

\section{Discussion and Conclusions} \label{sec:dis}

In this work, the SF analysis is used to study the correlations of RM variation for PSR B1744-24A and repeating FRBs. 
The perfect fit of SF for PSR B1744-24A demonstrates that binary systems should have a geometric component. The SF of RM for FRB 20201124A has a similar behavior as that of PSR B1744-24A, which suggests a binary origin of it. We have found strong evidence for the geometric component of SF for the first time. Besides, FRB 20180916B, FRB 20190520B and FRB 20121102A all have a potential geometric component in the SFs of RM.
The fitting results by the geometric component for these FRBs are shown in Table \ref{sffit}. 
Due to the limited data, the periodic feature in the geometric component caused by orbital motion has not been observed in these FRBs. So, the fitting results only give a lower limit of the orbital period and $\rm{RM_0}$. 

The orbital period of FRB 20201124A derived from geometric component fitting is likely to deviate from the true value for several reasons. First, an eccentric orbit is required to explain the steady phase of the RM of FRB 20201124A \citep{Wang2022} and Equation~(\ref{equ:2}) is only applicable for the circular orbit. The uneven sampling intervals of RM observations can lead to the SF being dominated by data from specific orbital phases (e.g. near the pericenter), causing the period inferred from the orbital velocities during these phases to be shorter than the overall orbital period. Second, the evolution of RM may not be strictly periodic, as it is influenced by clumps within the stellar disk \citep{Zhao2023}, which disrupt the ordered magnetic field \citep{Chernyakova2021, Chen2019}, as well as by interactions between the magnetar and its companion star. Therefore, the RM evolution in binary system is very likely to be non-periodic \citep{Li2023,Connors2002,Johnston1996,Johnston2001,Johnston2005}, making traditional methods for detecting periodicity, such as the Lomb-Scargle method and autocorrelation function (ACF), poorly suited for analyzing the RM of FRBs. In contrast, SF analysis is effective at identifying potential geometric modulations in the evolution of RM in FRBs.

In addition to the decline phase and periodic oscillations at larger temporal separations, the geometric component exhibits a power-law rising phase when $\Delta t < 2 \pi/\omega$. This behavior resembles the power-law feature of the SF of turbulence under the proper motion assumption. For FRBs that do not show a clear oscillatory phase in their SF (e.g., Figure \ref{fig4}), it is essential to differentiate between the periodic geometric and turbulent components by analyzing their respective power-law indices. Equation (\ref{equ:2}) can be reformulated in logarithmic space, yielding the first-order expression:
\begin{equation}
\rm{log}(D_g)= \rm{log}(\frac{1}{4}\mathrm{RM}_0^2(\omega \Delta t)^2) = C + 2\rm{log}(\Delta t),
\end{equation}
where C is a constant. The geometric component demonstrates a rising phase with a power-law index of 2, which differs from the expected indices listed in Table \ref{tab:PW}. However, as previously mentioned, both power-law fitting and geometric fitting are subject to considerable uncertainties due to the limited number of data points. Determining whether these FRBs exhibit geometric components or turbulent component will require further investigation in the future.
 
The power-law indices of RM SFs for FRB 20190303A and FRB 20180301A are weakly sensitive to how the data are binned. The different bins have only a little effect on the results. The range of variation of the power indices $m$ for different methods of binning is relatively small (e.g. $|\Delta m|\lesssim0.2$). The power-law indices are consistent with the expectation of the two-dimensional Kolmogorov turbulent medium \citep{Minter1996}.
 
The SFs of DM and RM for these FRBs are significantly distinct. The DM SFs for these FRBs are almost unchanging across all temporal separations. The spectral indices for SF of DMs are from $-0.09$ to $0.09$, which demonstrates that the SF of DM is independent of temporal separation. The stochastic process may be the intrinsic mechanism driving the DM evolution. This is consistent with the irregular and stochastic evolution of the DM time series of these FRBs. It can be inferred that the changes in electron density of these FRBs in the observing time are both tiny. Combined with the RM variation, it suggests that the parallel component of the magnetic field is dynamically evolving with time around these FRB sources. 

Compared with the complex binary model, which involves many unknown parameters to interpret RM variations, SF analysis has the advantage that the geometric shape of the geometric component is fixed. In our fitting, it is difficult to force-fit the observational data through fine-tuning the parameters, as there are only two free parameters available to adjust its specific positioning. Therefore, it can give a much more reliable assessment of the binary origin of FRB 20201124A. However, SF analysis heavily depends on the availability of sufficient RM data. For FRBs with limited observational data, the reliability of SF results is significantly lower compared to those derived from FRBs with abundant data (e.g. FRB 20201124A). In the future, monitoring repeating FRBs with radio telescopes will yield larger RM samples, enabling a more reliable diagnosis of their environments and physical origins. 

\section*{Acknowledgments}
We thank the anonymous referee for helpful comments. We thank Y.P. Yang, Y. Wu and R. Mckinven for helpful comments and discussions, Y.Q. Wang for drawing Figure \ref{fig1}, and J.C. Jiang for providing the RM data. This work was supported by the National Natural Science Foundation of China (grant Nos. 12494575 and 12273009), the National
SKA Program of China (grant No. 2022SKA0130100) and Postgraduate Research \& Practice Innovation Program of Jiangsu Province (KYCX24\_0184). This work made use of data from Five-hundred-meter Aperture Spherical radio Telescope (FAST), a Chinese national mega-science facility built and operated by the National Astronomical Observatories, Chinese Academy of Sciences.

\bibliographystyle{aasjournal}
\bibliography{ref1}

\end{document}